\documentstyle[12pt,aaspp4]{article}
\begin{document}

\title{CTQ 414: A New Gravitational Lens \altaffilmark{1}}

\author{
 Nicholas D. Morgan\altaffilmark{2},
 Alan Dressler\altaffilmark{3},
 Jos\'{e} Maza\altaffilmark{4,5},
 Paul L. Schechter\altaffilmark{2}, and
 Joshua N. Winn\altaffilmark{2}
}

\altaffiltext{1}{Based on observations carried out at the Cerro Tololo
 Interamerican Observatory (CTIO), the Las Campanas Observatory (LCO),
 and the National Radio Astronomy Observatory (NRAO) Very Large Array
 (VLA).  CTIO is part of the National Optical Astronomy Observatories,
 which are operated by the Association of Universities for Research in
 Astronomy, Inc., under cooperative agreement with the National Science
 Foundation.  The NRAO is a facility of the National Science Foundation
 operated under cooperative agreement by Associated Universities, Inc.}

\altaffiltext{2}{Department of Physics, Massachusetts Institute of
 Technology, Cambridge MA 02139; ndmorgan@mit.edu, schech@achernar.mit.edu,
 jnwinn@mit.edu}

\altaffiltext{3}{Carnegie Observatories, 813 Santa Barbara Street,
 Pasadena, CA 91101; dressler@ociw.edu}

\altaffiltext{4}{Departamento de Astronom\'{\i}a, Universidad de Chile,
 Casilla 36-D, Santiago, Chile; jmaza@das.uchile.cl}

\altaffiltext{5}{C\'{a}tedra Presidencial de Ciencias 1996-1998}

\begin{abstract}
We report the discovery and ground based observations of the new 
gravitational lens CTQ 414.  The source quasar lies at a redshift of 
{\em z} = 1.29 with a {\em B} magnitude of 17.6.  Ground based optical 
imaging reveals two point sources separated by 1\farcs2 with a magnitude 
difference of roughly 1 mag.  Subtraction of two stellar point spread 
functions from images obtained in subarcsecond seeing consistently leaves 
behind a faint, residual object.  Fits for two point sources plus an 
extended object places the fainter object collinear with the two brighter 
components.  Subsequent HST/NICMOS observations have confirmed the 
identification of the fainter object as the lensing galaxy.  VLA 
observations at 8.46 GHz reveal that all components of the lensing system 
are radio quiet down to the 0.2 mJy flux level.
\end{abstract}


\section{INTRODUCTION}

Gravitational lensing is a powerful tool with a wide range of
astrophysical applications (Kochanek and Hewitt 1996).  In particular,
multiply imaged quasars have become a useful probe for a number of
cosmological investigations, such as measurements of the Hubble constant
(Refsdal 1964) and statistical constraints on the cosmological constant
(Kochanek 1996).  The usefulness of gravitationally lensed quasars,
however, is limited by the relatively few systems discovered to date.
In this paper we report our discovery and analysis of a new multiply 
imaged quasar.

CTQ 414 (1$^{\scriptsize{\mbox{h}}}$ 58$^{\scriptsize\mbox{m}}$
41$\fs$43, -43$\arcdeg$ 25$\arcmin$ 3$\farcs$4, J2000.0) was originally
identified as a {\em z} = 1.29, {\em B} = 17.2 quasar from the Cal\'{a}n
Tololo Survey (CTS) (Maza {\em et al.} 1995), a survey designed to
discover quasars and emission-line galaxies in the southern hemisphere.
In August of 1997, optical observations of approximately 200 of the CTS
quasars were carried out at the Cerro Tololo Interamerican Observatory
(CTIO).  The primary purpose of this effort was to determine if any of
the selected CTS quasars exhibited evidence of arcsecond, multiple images
that would arise from gravitational lensing.  Prior to this run, another
Cal\'{a}n Tololo quasar, CTQ 286, had been found to be lensed by
Claeskens {\em et al.} (1996).  CCD exposures of CTQ 414 immediately
revealed it to be double, with a separation of 1\farcs2 evident in
{\em B, V, R,} and {\em I} filters.

We present the details of these observations, along with their
subsequent analysis, in \S2.  Also described in \S2
are follow-up observations conducted at the Las Campanas Observatory (LCO)
two weeks later, as well as our photometric analysis of the lens
components utilizing astrometric positions obtained from HST/NICMOS 
imaging.  In \S3 we present our astrometric and photometric results
for field stars in the quasar field.  In May of 1998, the Very Large 
Array (VLA) was used to search for radio counterparts of CTQ 414.  We 
discuss these observations in \S4.  Finally, in \S5 we summarize our 
findings for CTQ 414.

\section{OPTICAL OBSERVATIONS AND ANALYSIS}
\subsection{Initial Optical Imaging: Detection of system duplicity}

Initial optical observations of CTQ 414 were obtained by one of
us (P.L.S.) at CTIO between the nights of 1997 August 26-30.  The 1.5 m
telescope equipped with the Tektronix 2048 No. 6 CCD was used, although
only the central 1536x1536 array was read out.  The field of view of the
camera was 6.2 arcmin square with a scale of 0.2417 arcsec/pixel, a gain
of 2 e$^-$/ADU, and a read noise of 5 e$^-$.  A total of 20 exposures of
CTQ 414 were obtained with Johnson {\em BV} and Kron-Cousins {\em RI} 
filters on the nights of the $26^{th}$, $27^{th}$, and $29^{th}$.  Each 
exposure lasted 300 s, and were obtained through airmasses ranging from
1.027 to 1.143 over the course of the three nights.
Seeing conditions ranged from 1\farcs16 to 1\farcs93 FWHM.  Multiple
60 s {\em BVRI} exposures of two Landolt photometric standard fields
(fields PG0231+051 and PG2331+055) were also obtained during two of the
above nights.  Our exposures of PG0231+051 and PG2331+055 each contained 
five and three Landolt standard stars, respectively.  A log of the 
observations for CTQ 414 are presented in Table 1.  Figure 1 shows an 
{\em R} band exposure of CTQ 414 and the surrounding field obtained in
1$\farcs25$ FWHM seeing.  The reader will note the relative absence
of comparably bright stars in this high latitude field.

Each CCD frame was bias-subtracted, trimmed, and flat-field corrected
using the Vista reduction program.  The flat-field images were cleaned
of cosmic rays using ``autoclean,'' a program written and kindly supplied
by J. Tonry.  As mentioned in \S1, images of CTQ 414 appeared double
in all exposures (see Figure 2a, note caption).  We therefore fit
the double images with two empirical point spread functions (PSFs)
using a variant of the program DoPHOT (Schechter {\em et al.} 1993),
designed to deal with close, point-like and extended objects (Schechter
and Moore 1993).  Star \#7 as shown in Figure 1 was used as the empirical 
PSF.  These fits yielded average separation distance between the two components
in $B$, $V$, $R$, and $I$ of 1\farcs200 $\pm$ 0\farcs006, 1\farcs222 $\pm$ 
0\farcs012, 1\farcs215 $\pm$ 0\farcs009, and 1\farcs193 $\pm$ 0\farcs010, 
respectively.  We note that the smaller separation in $I$ band is consistent 
with the presence of a relatively red object between the double image. 
However, no consistent or suggestive pattern of residuals were found upon 
fitting and subtracting the two PSFs, although the difficulty of fitting 
for multiple point-sources separated by less than a seeing disc should be noted.

\subsection{Followup Optical Imaging: Detection of a Third Object}

Following the initial imaging of CTQ 414 taken in August at CTIO, an
additional 12 exposures were obtained by one of us (A.D.) at LCO on the
night of 1997 September 10.  The du Pont 2.5 m telescope equipped with a
Tektronix No. 2 CCD detector was used.  The CCD camera has a field of
view of 3.9 arcmin square with a scale of 0.2278 arcsec/pixel.  The camera
was set in the \#3 gain position, providing a gain of 3.9 e$^-$/ADU and
a readnoise of 9.4 e$^-$.  Seeing for the night ranged from 0\farcs89 to
1\farcs08 FWHM, significantly better than the initial images from CTIO taken 
two weeks earlier.  All 12 images were taken in {\em R} band and each
exposure lasted 300 s.

The frames were bias-subtracted, trimmed, and flat-field corrected,
following the same procedure as mentioned above.  Simultaneous fitting
of two empirical PSFs (using the identical star as noted in the previous 
subsection) resulted in an average separation between the two components 
of 1\farcs198 $\pm$ 0\farcs001.  This separation agrees well with the 
average {\em R} band separation obtained from the CTIO data.  

Subtraction of the two empirical PSFs produced a consistent
pattern of residuals present in all 12 frames (see Figure 2b).  This pattern
consisted of a bright spot located slightly west of component A and northeast 
of component B, as well as two crescent-like arcs of positive 
residuals to the southeast of each component, and two regions of
negative ``cavities'' located at the centers of A and B.  The amplitude
of the bright spot is quite small, of order 2-3\% of the central intensity 
of component A.  The crescent is 1-2\% of A's central intensity, while the 
cavities for A and B were 3-4\% and 2-3\% of A's central intensity, 
respectively.  

One possible explanation for this pattern of residuals is that the
fitting program is trying to account for three sources of light with just
two point sources.  If the unaccounted source of light was located between
and displaced slightly to the northwest from the two components, then the PSFs
would be dragged off-target from the centers of components A and B in an 
attempt to cover the extra light.  This would produce the observed crescent 
pattern of residuals along the southeast edge.  Also, the observed cavities 
would arise from attributing excessive flux to components A and B in an attempt 
to account for the three sources of light in the system.

Another possibility is that we are seeing an artifact of a poor
PSF template.  Such an effect might arise from variations in the PSF across
the surface of the CCD, or from mistakenly using an extended object as our model 
point source.  However, we do not believe that either of these explanations is 
correct, since the identical residual pattern persists even when different 
stars from across the face of the detector are used as our PSF.  

Following the suggestive pattern of the residuals, we decided to fit each
frame of the LCO data for two point sources plus an extended object
(hereafter ``C''), which we interpret here as the lensing galaxy.  The object was 
modeled as a circularly symmetric pseudogaussian as descirbed in 
Schechter {\it et al.} (1993), convolved with the seeing conditions.  The positions 
and fluxes of all three components, as well as the size $\sigma$ of the pseudogaussian, 
were treated as free parameters.  

The averaged results for these fits placed component C collinear with
components A and B and at an angular distance of 0$\farcs$72 $\pm$ 0$\farcs$03 
away from A.  The separation between components A and B was determined as 
1$\farcs$29 $\pm$ 0$\farcs$02.  Note that this solution for the lensing galaxy 
places it roughly midway between components A and B.  The rms scatter in the galaxy 
position was 0\farcs034 along the N-S direction, and 0\farcs088 along the E-W 
direction, in both cases small compared to the relative distances between the 
three components.  Flux ratios of components B to A and components C to A were 
32.6\% $\pm$ 0.6\% and 20.5\% $\pm$ 0.5\%, respectively.  The average $\sigma$ 
of the pseudogaussian used to model component C was found to be 0\farcs32 $\pm$ 
0$\farcs$010.

This result for the position of the lensing galaxy, that it is very
nearly equidistant from the two quasar images, would appear to be
inconsistent with the unequal flux ratio of the two quasar components.
For an isothermal lens, the flux ratio for the components would be
proportional to the ratio of their distances from the lensing galaxy.

A very similar situation was encountered in the case of FBQ 0951+2635
(Schechter {\it et al.} 1998), with the position of the lensing galaxy too
well centered for the unequal flux ratio.  The parameters for the
ground based observations for this system were quite similar to those
for CTQ 414.  Subsequent imaging of FBQ 0951+2635 with HST/NICMOS confirmed
the presence of the lensing galaxy (McLeod {\it et al.} 1998), but 
placed it considerably closer to the fainter image than did the ground 
based observations.  A similar systematic error of this sort with CTQ 414 
would not be surprising, given the difficulty of measuring a third object between 
two objects separated by only one seeing disk.  

\subsection{Confirmation of the Lensing Galaxy}

In an attempt to confirm the lensing hypothesis, CTQ 414 was placed onto 
the CfA-Arizona Space Telescope Lens Survey (CASTLES) (Kochanek {\it et al.} 1998) 
observing program at the suggestion of the authors.  The suspected lens was imaged 
by HST/NICMOS in four 640 s H band exposures on 1998 August 4.  These exposures clearly 
reveal the extended emission of the lensing galaxy, and place it approximately collinear 
with the two quasar images (Falco {\it et al.} 1998).  Gaussian fits for the positions of 
components B and C with respect to component A places B 1\farcs22 from A at 
a PA of 251\fdg0 E of N, and the nucleus of component C 0\farcs80 from A
at a PA of 253\fdg4 E of N.  As suspected, the true position of the 
lensing galaxy is $\sim$10\% closer to the fainter quasar image than indicated 
by the ground-based data.  The HST images of CTQ 414 is available to 
download from the CASTLES ftp site, or may be viewed online at the CASTLES
homepage at http://cfa-www.harvard.edu/glensdata. 

\subsection{Photometric Analysis of Lens Components}

Photometry of the lens components was performed using the same variant of
DoPHOT described above.  Photometric solutions to the LCO 
data set were obtained by fixing the relative separations of components A, B, 
and C at the corresponding HST positions.  The fluxes of the 
three components, as well as the overall position of the system and the size 
$\sigma$ of the lensing galaxy, were treated as free parameters.  In the 
process of fixing the relative separations of the system components, appropriate 
steps were taken to account for the plate scale and chip orientation of the 
LCO detector.  Table 2 summarizes our photometric results for the LCO data set.
Here we present the separate magnitudes of components A, B, and C, as well as
magnitude differences between system components and the combined magnitude from
the two quasar images.

A similar attempt was made to determine the colors of the lens 
components using the $BVRI$ data from CTIO.  Again, relative separations 
were fixed at the corresponding HST values, while the overall 
position of the system, the fluxes of components A, B, and C, and the 
size $\sigma$ of the lensing galaxy were treated as free parameters.  
Appropriate steps were again taken to account for the plate scale and chip 
orientation of the CTIO detector.  Unfortunately, treating the flux and size of
component C as free parameters introduced too much freedom into these models, 
and the fits failed to converge to our satisfaction.  Fixing the size $\sigma$ 
of the lensing galaxy across all wavelengths at the LCO $R$ band value resulted 
in no noticeable improvement.  The quality of seeing on the CTIO 
data set is simply not good enough to perform stable photometry of component C.  
Reliable colors of all three system components will require better images.

In order to circumvent this problem to some extent, we decided to set the flux 
of the lensing galaxy to zero and only solve for the fluxes of components A and B.  
Relative positions were still held constant at the corresponding HST values, 
while the overall position of the system was free to vary.  Table 3 presents
our magnitude results for the combined flux of components A and B.
Admittingly, these results contain a systematic error by failing to take into 
account light from the lensing galaxy.  Upon comparison with the corresponding 
LCO result presented in Table 2, we find that the combined CTIO A+B magnitude in 
{\em R} is brighter than the corresponding LCO result by 0.08 mag.  Under the 
assumption that the lensing galaxy gets brighter in redder wavelengths, we would 
expect a correspondingly smaller systematic error in {\em B} and {\em V} bands, and a 
larger one in {\em I}.  It is hoped that these results will provide a handle 
on future variability of the system across {\em BVRI} wavelengths.

\section{PHOTOMETRY AND ASTROMETRY STANDARDS}

\subsection{Photometric Standards}

Photometric and astrometric results for stars in the CTQ 414 field were 
obtained for use with future observations.  Eight nearby reference stars 
within a 4 arcmin radius from the target lens were chosen for this purpose.
These stars are identified by the labels shown in Figure 1.   

Observations of the Landolt (1992) standard fields PG0231+051 (PG0231+051, 
PG0231+051A, PG0231+051B, PG0231+051C, PG0231+051D) and PG2331+055 
(PG2331+055, PG2331+055A, PG2331+055B) were used to derive color terms and 
zero-point offsets for calibration onto the Johnson-Kron-Cousins photometric 
system.  Table 4 lists the color terms used for the transformations as well 
as the number of standard stars $N$ used and the corresponding 
rms scatter of the fit.  Color terms were extracted from the 
PG0231+051 field, making use of the field's greater sampling of $B-V$ indices as
compared to the PG2331+055 field.  Zero-point offsets were derived from the 
PG2331+055 field.  In deriving the $I$ band color term, the faintest of the five 
observed PG0231+051 standard stars was discarded (see below).  
We have solved for the transformation equations in the form 
\begin{equation} X - x = const. + a_1 * (B - V)\end{equation}
\begin{equation} B - V = const. + a_2 * (b - v)\end{equation}
where {\em X} represents the apparent magnitude in the standard {\em BVRI}
system, {\em x} the extinction corrected instrumental magnitude above the earth's 
atmosphere, and \begin{math} a_1 \end{math} and \begin{math} a_2 \end{math} 
the respective color terms for the $X-x$ and $B-V$ transformations.
Corrections for atmospheric extinction were applied using ``typical'' 
extinction coefficients as cited in the 1990 CTIO Facilities Manual 
(\begin{math} k_B = 0.22, k_V = 0.11, k_R = 0.08, k_I = 0.04 \end{math}).

In the process of reducing the standard fields, the relatively faint star PG0231+051 
(Landolt values of $I$=16.64 and $B-V$=-0.33) was discovered to be brighter in {\em I} by
0.30 $\pm$ 0.05 mag with respect to Landolt's (1992) listed magnitude.  This residual 
is in the sense $I_{stnd} - I_{obs}$, where $I_{stnd}$ is the value 
reported by Landolt (1992), and $I_{obs}$ has been computed from the 
transformations given above.  Similar $I$ band discrepancies for PG0231+051, as 
compared to the Landolt (1992) standard value, have been reported by Geisler (1996), 
who found an $I$ residual of 0.23 mag, and also by Rosvick (1995), who found 
an $I$ residual of 0.15 mag.  The residual reported by Geisler is in the same
sense as mentioned above, while Rosvick does not report the sense of his 
residual.  Because of the observed discrepancy, the $I$ band observations of PG0231+051 
were not included in the solution of the $I$ band color term.

The empirical PSF fitting described in the previous section yields magnitudes 
of stars in the quasar field with respect to the template PSF star.  With the template 
star placed onto the standard photometric system, magnitudes for 
the field stars are straightforward to obtain.  Apparent
$BVRI$ magnitudes for our eight reference stars are listed in Table 5,
along with respective standard errors of the mean ($\sigma/N^{1/2}$) as derived from
frame-to-frame scatter.  These error bars do not include uncertainties in the reference 
star's calibration, which are listed in the footnote to Table 5.
Object numbers in Table 5 correspond to the labels shown in Figure 1.  These 
local standards were used to calibrate the LCO photometry results reported in Table 2.

It should be noted that images taken from CTIO are afflicted by coma aberration.  
The CTIO 1.5 m is a Ritchey-Chr\'{e}tien telescope designed to operate free
from coma aberration at an $f$ ratio of $f/7.5$, but not at $f/13.5$.  
Observations at CTIO were carried out at $f/13.5$ to make use of the smaller
pixel scale at that $f$ ratio.  The presence of coma 
aberration results in a slight, off-axis distortion in the PSF shape across the face of 
the detector.  By comparing magnitudes derived from multiple PSF fitting, we estimate that 
coma aberration for our images is a small effect, introducing uncertainties in our magnitude 
determinations of 0.02 mag for stars on the largest contours of constant wavefront aberration.  

\subsection{Astrometric Standards}

Astrometric solutions were also obtained for the eight reference stars in the quasar 
field using one of the {\em R} band exposures taken through a seeing of 1\farcs32 FWHM.  
This particular exposure was chosen since it yielded the lowest rms position 
errors with respect to standard coordinates.  Standard coordinates for the
objects were taken from the APM Sky Catalogue at Cambridge, England.
The astrometric results are included in Table 5.  The reported positions are in 
the form of offsets from star \#7, in the sense of star {\em x} minus star \#7.  

\section{RADIO OBSERVATIONS}

On 1998 May 19 we searched for radio emission from CTQ 414, using the NRAO Very Large 
Array (VLA).  The search was carried out at 8.46 GHz, while the VLA was in the A
configuration.  This provided an east-west resolution of 0$\farcs$2.  Our
14-minute observation bracketed the transit time of CTQ 414, at an altitude
of 12 degrees, permitting a north-south resolution of approximately 1$\arcsec$.

No significant sources of radio flux were detected within a 5$\arcsec$ error 
circle around the position of CTQ 414.  The rms noise level in
this field was 0.065 mJy per synthesized beam, so our observation rules out
(at the 3$\sigma$ level) any sources of compact flux above 0.2 mJy.
This is unfortunate but not particularly surprising, since the large majority
of known quasars are radio quiet.  

\section{SUMMARY AND CONCLUSIONS}

	We have reported our discovery and photometric analysis of the new gravitational 
lens CTQ 414.  Ground-based optical images of the quasar appear double with a separation of 
1$\farcs$2 and a magnitude difference between the quasar images of roughly 1 mag.
Fitting and subtracting two empirical point spread functions to 
images obtained in subarcsecond seeing consistently leaves behind a 
faint, residual object.  Fits for two PSFs plus an extended object place the 
extended object collinear with the pair of brighter components.  Subsequent HST imaging
with NICMOS has indeed confirmed the extended object as the lensing galaxy.  
We have shown that ground-based photometric analysis of all three lens components is 
feasible with subarcsecond seeing conditions, and hope that the photometric analysis 
presented in this paper will provide a handle on any future variability of the quasar images.

\acknowledgements 

	N.D.M. and P.L.S. gratefully acknowledge the support of the U.S. 
National Science Foundation through grant AST96-16866.  J.N.W. thanks the 
Fannie and John Hertz Foundation for financial support.  J.M. thanks 
FONDECYT, Chile, for support through grant 1980172.

\clearpage
\figcaption[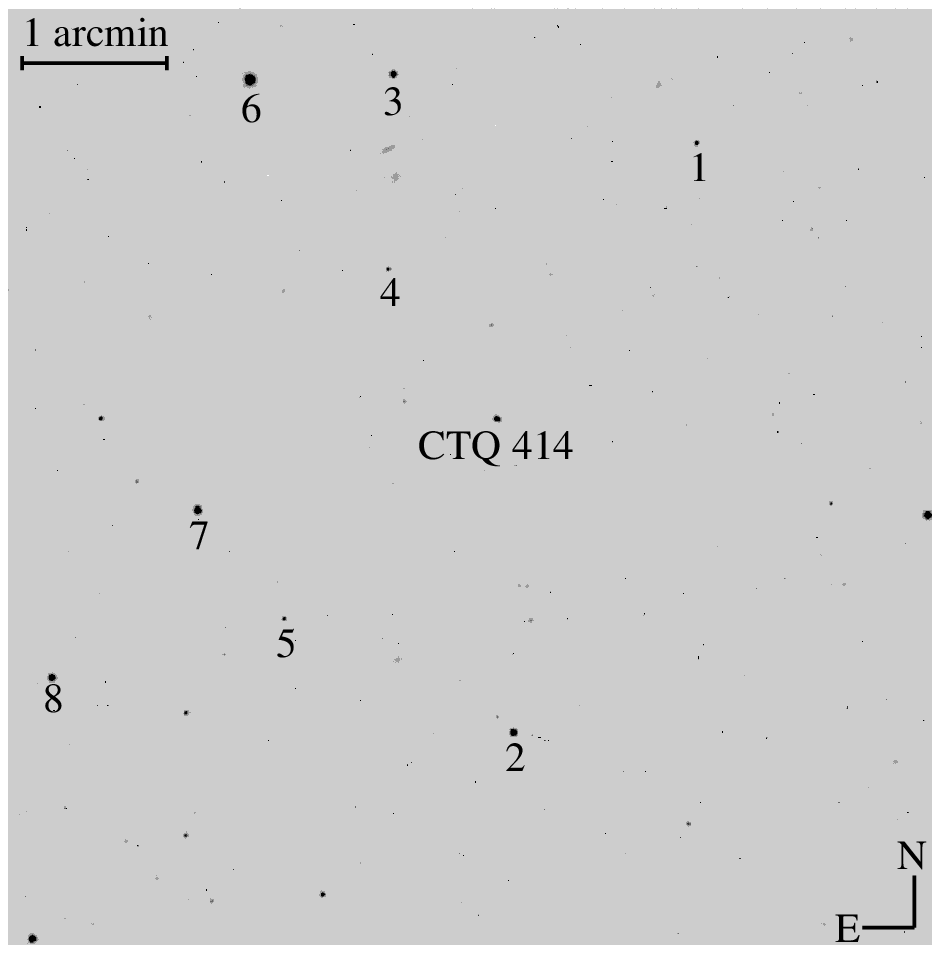] 
	{ a) A 300 s {\em R} band image of CTQ 414 and surrounding stars 
taken at CTIO.  Photometric and astrometric results have been obtained for 
the stars labeled 1 through 8.  Labels have been placed directly beneath 
the objects they identify.  \label{fig1}}

\figcaption[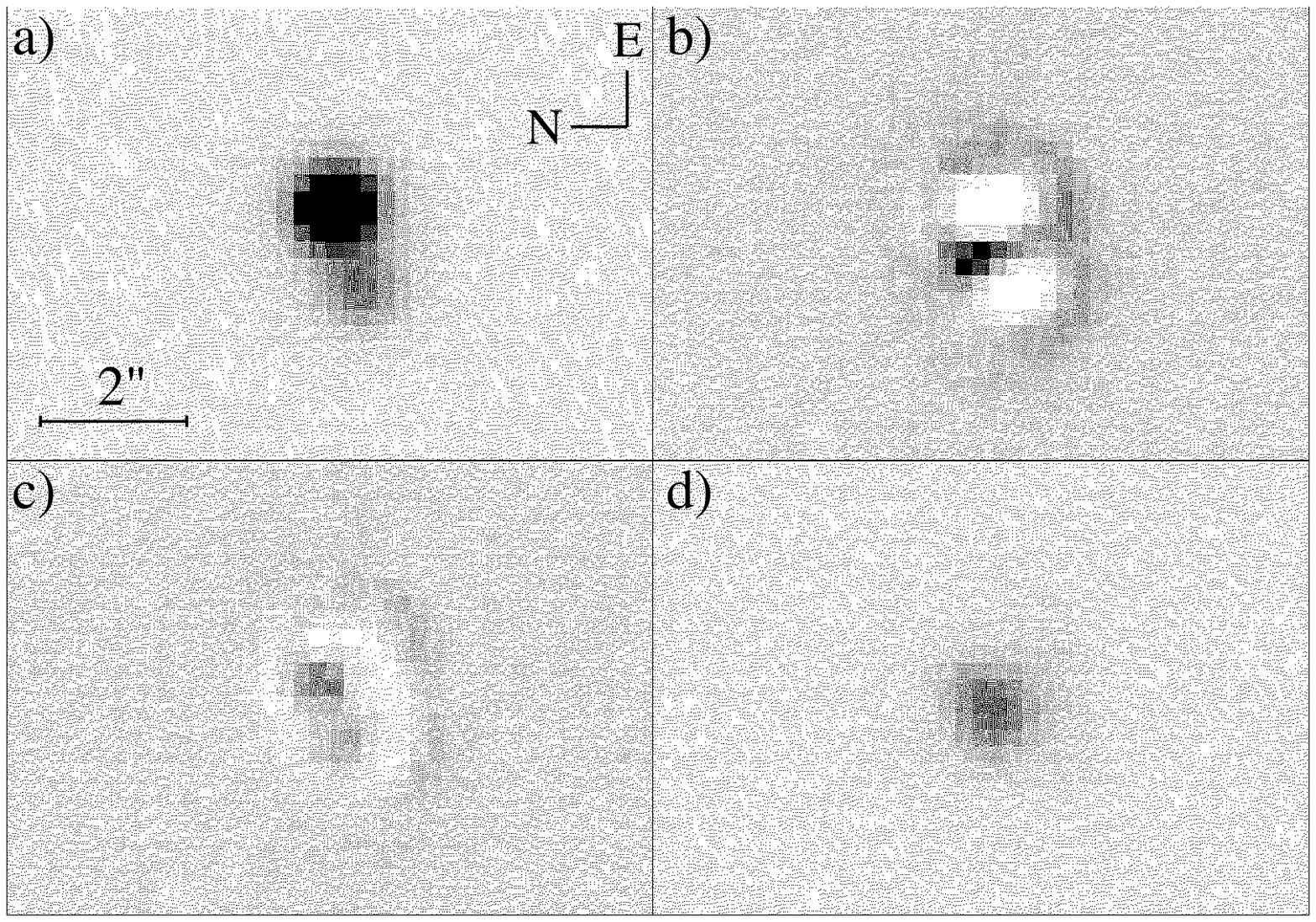]
	{ a) A summed image of the 12 {\em R} band exposures of CTQ 414
taken at LCO.  The scale for the image is 0$\farcs$2278 per pixel.  East 
is up and North is to the left.  The fainter component is at PA 
251$^\circ$ E of N.  b)  Summed image after fitting and subtracting two 
stellar point spread functions from the individual frames.  c) Summed 
image after fitting and subtracting two stellar point spread functions 
plus a pseudogaussian from the individual frames.  d) Summed image 
after fitting and subtracting two stellar point spread functions from 
the individual frames, but leaving the pseudogaussian unsubtracted.  
Panels b) and c) are displayed at a factor of 12 higher contrast than 
panel a).  Panel d) is at a factor of 4 higher contrast than panel a).
\label{fig2}}

\clearpage




 

\makeatletter
\def\jnl@aj{AJ}
\ifx\revtex@jnl\jnl@aj\let\tablebreak=\nl\fi
\makeatother


\begin{deluxetable}{ccccccc}
\tablecaption{Log of Observations for CTQ 414 at CTIO
\label{TABLE1}}
\tablenum{1}
\tablewidth{0pt}
\tablehead{
\colhead {Frame \#} &
\colhead {Filter} &
\colhead {FWHM (\arcsec)} &
\colhead { } &
\colhead {Frame \#} &
\colhead {Filter} &
\colhead {FWHM (\arcsec)}}
\startdata
104 & R & 1.31 & & 233 & R & 1.30\nl
105 & B & 1.48 & & 399 & R & 1.42\nl
106 & V & 1.43 & & 401 & R & 1.25\nl
107 & B & 1.38 & & 402 & B & 1.18\nl
108 & I & 1.30 & & 403 & V & 1.16\nl
211 & R & 1.32 & & 405 & I & 1.30\nl
212 & B & 1.47 & & 406 & I & 1.26\nl
213 & V & 1.93 & & 408 & I & 1.24\nl
214 & V & 1.59 & & 409 & B & 1.39\nl
215 & I & 1.47 & & 410 & V & 1.25\nl
\enddata
\end{deluxetable}


\clearpage



 

\makeatletter
\def\jnl@aj{AJ}
\ifx\revtex@jnl\jnl@aj\let\tablebreak=\nl\fi
\makeatother


\begin{deluxetable}{ccccc}
\tablecaption{Photometric Solutions for LCO Data
\label{TABLE2}}
\tablenum{2}
\tablewidth{0pt}
\tablehead{
\colhead {} &
\colhead {$\Delta$ RA (s)} &
\colhead {$\Delta$ Dec ($\arcsec$)} &
\colhead {$\Delta$m$_R$} &
\colhead {m$_R$}}
\startdata
A   &  0.                   &   0.                 &   0.                 & 17.338 $\pm$ 0.001 \nl
B   & -0.1061               &  -0.398              &  -1.149 $\pm$ 0.004  & 18.487 $\pm$ 0.004 \nl
C   & -0.0700               &  -0.228              &  -2.102 $\pm$ 0.027  & 19.440 $\pm$ 0.028 \nl
A+B &  ---                  &  ---                 &   ---                & 17.018 $\pm$ 0.003 \nl
\enddata
\tablecomments{Relative positions for components A, B, and C were 
obtained from HST/NICMOS imaging and were held fixed during our
photometric solution.  Error bars are from the observed dispersion
between the images and do not include uncertainties in the magnitude
of the PSF template star (see footnote to Table 5).}
\end{deluxetable}


\clearpage



 

\makeatletter
\def\jnl@aj{AJ}
\ifx\revtex@jnl\jnl@aj\let\tablebreak=\nl\fi
\makeatother


\begin{deluxetable}{ccccc}
\tablecaption{Photometric Solutions for CTIO Data
\label{TABLE3}}
\tablenum{3}
\tablewidth{0pt}
\tablehead{
\colhead{} &
\colhead{B} &
\colhead{V} &
\colhead{R} &
\colhead{I}}
\startdata
m$_{A+B}$& 17.627 $\pm$ 0.004& 17.334 $\pm$ 0.010& 16.932 $\pm$ 0.007& 16.728 $\pm$ 0.004 \nl
\enddata
\tablecomments{Error bars are from the observed dispersion
between the images and do not include uncertainties in the magnitude
of the PSF template star (see footnote to Table 5).}
\end{deluxetable}

\clearpage



 

\makeatletter
\def\jnl@aj{AJ}
\ifx\revtex@jnl\jnl@aj\let\tablebreak=\nl\fi
\makeatother


\begin{deluxetable}{ccccc}
\tablecaption{Color Terms for Transformation Equations
\label{TABLE4}}
\tablenum{4}
\tablewidth{0pt}
\tablehead{
\colhead{Index} &
\colhead{N} &
\colhead{$a_1$} &
\colhead{$a_2$} &
\colhead{rms}}
\startdata
B  & 15& -    0.0807& --    & 0.0099 \nl
V  & 10& \phs 0.0110& --    & 0.0244 \nl
R  & 10& -    0.0194& --    & 0.0134 \nl
I  &  8& \phs 0.0363& --    & 0.0184 \nl
B-V& --& --         &-0.0847& 0.0218 \nl
\enddata
\end{deluxetable}


\clearpage



 

\makeatletter
\def\jnl@aj{AJ}
\ifx\revtex@jnl\jnl@aj\let\tablebreak=\nl\fi
\makeatother


\begin{deluxetable}{crrrrcccc}
\tablecaption{Relative Astrometry and Absolute Photometry for Nearby Reference Stars
\label{TABLE5}}
\tablenum{5}
\tablewidth{0pt}
\tablehead{
\colhead{Object} &
\colhead{$\Delta \alpha (^s)$ } & 
\colhead{$\Delta \delta (\arcsec)$ } &
\colhead{$m_B$} &
\colhead{$m_V$} &
\colhead{$m_R$} &
\colhead{$m_I$}}
\startdata
1&  -18.255&  149.32& 19.563 $\pm$ 0.018 & 18.519 $\pm$ 0.023 & 17.987 $\pm$ 0.008 & 17.482 $\pm$ 0.025 \nl
2&  -11.794&  -87.49& 19.462 $\pm$ 0.009 & 17.831 $\pm$ 0.009 & 16.775 $\pm$ 0.003 & 15.429 $\pm$ 0.007 \nl
3&   -7.023&  175.32& 17.536 $\pm$ 0.017 & 16.812 $\pm$ 0.014 & 16.450 $\pm$ 0.001 & 16.077 $\pm$ 0.002 \nl
4&   -6.938&   97.31& 21.162 $\pm$ 0.020 & 19.709 $\pm$ 0.015 & 18.899 $\pm$ 0.006 & 18.090 $\pm$ 0.011 \nl
5&   -3.270&  -42.97& 19.332 $\pm$ 0.007 & 18.822 $\pm$ 0.006 & 18.514 $\pm$ 0.008 & 18.189 $\pm$ 0.007 \nl
6&   -1.733&  172.50& 15.598 $\pm$ 0.009 & 14.789 $\pm$ 0.020 & 14.373 $\pm$ 0.003 & 13.957 $\pm$ 0.001 \nl
7&    0.000&    0.00& 16.472 $\pm$ 0.001 & 16.121 $\pm$ 0.001 & 15.904 $\pm$ 0.001 & 15.671 $\pm$ 0.001 \nl
8&    5.302&  -67.74& 17.323 $\pm$ 0.016 & 16.697 $\pm$ 0.004 & 16.388 $\pm$ 0.014 & 16.034 $\pm$ 0.008 \nl
\enddata
\tablecomments{
Magnitudes have been derived with respect to the PSF template star (\#7).  Reported error bars are from
the observed dispersion between the images and do not include uncertainties in the magnitude of the reference 
star.  $BVRI$ uncertainties in the reference star's magnitude are $\pm$0.007, $\pm$0.011, $\pm$0.014, and 
$\pm$0.016 mag, respectively, and must be added in quadrature to the error bars quoted above.}
\end{deluxetable}


\clearpage

\begin{figure}[h]
\vspace{7.0 truein}
\includegraphics{NickMorgan.fig1.ps}
\end{figure}

\clearpage

\begin{figure}[h]
\vspace{7.0 truein}
\includegraphics{NickMorgan.fig2.ps}
\end{figure}

\end{document}